\documentclass{PoS}

\usepackage{amsmath}

\title{
Tempered Lefschetz thimble method and its application 
to the Hubbard model away from half filling%
\footnote{Report No.: KUNS-2777}
}

\ShortTitle{
Tempered Lefschetz thimble method and the Hubbard model}

\author{\speaker{Masafumi Fukuma},$^a$ Nobuyuki Matsumoto$^{a}$
and Naoya Umeda$^{b}$\\
\llap{$^a$}Department of Physics, Kyoto University\\
Kyoto 606-8502, Japan\\
\llap{$^b$}PricewaterhouseCoopers Aarata LLC\\
Otemachi Park Building, 1-1-1 Otemachi, Chiyoda-ku, Tokyo 100-0004, Japan\\
E-mail: \email{fukuma@gauge.scphys.kyoto-u.ac.jp}, 
\email{nobu.m@gauge.scphys.kyoto-u.ac.jp},
\email{naoya.umeda1134@gmail.com}}

\abstract{
The tempered Lefschetz thimble method (TLTM) 
is a parallel-tempering algorithm 
towards solving the numerical sign problem. 
It tames both the sign and ergodicity problems simultaneously 
by tempering the system with the flow time of continuous deformations 
of the integration region. 
In this article, 
after reviewing the basics of the TLTM, 
we explain a new algorithm within the TLTM 
that enables us to estimate the expectation values precisely 
with a criterion ensuring global equilibrium 
and the sufficiency of the sample size. 
To demonstrate the effectiveness of the algorithm, 
we apply the TLTM to the quantum Monte Carlo simulation 
of the Hubbard model away from half filling 
on a two-dimensional lattice of small size, 
and show that the obtained numerical results agree nicely with exact values.
}

\FullConference{37th International Symposium 
on Lattice Field Theory - Lattice2019\\
16-22 June 2019\\
Wuhan, China}

\newcommand{\bbC}{{\mathbb{C}}}
\newcommand{\bbR}{{\mathbb{R}}}

\newcommand{\bx}{{\textbf{x}}}
\newcommand{\by}{{\textbf{y}}}

\begin{document}

\section{Introduction
\label{sec:introduction}}
The sign problem is one of the major obstacles 
for numerical calculations in various fields of physics, 
including 
finite density QCD \cite{Aarts:2015tyj}, 
quantum Monte Carlo (QMC) calculations of quantum statistical systems 
\cite{Pollet:2012}, 
and the numerical simulations of real-time quantum field theories. 
In this article, we argue that a new algorithm 
which we call the ``tempered Lefschetz thimble method'' (TLTM) 
\cite{Fukuma:2017fjq,Fukuma:2019wbv}
is a promising method towards solving the sign problem. 
We exemplify its effectiveness 
by applying the method  
to the QMC simulations of the Hubbard model away from half filling 
on a two-dimensional lattice of small size 
(see \cite{Fukuma:2019wbv} for a detailed analysis).%
\footnote{
The application of Lefschetz thimble methods to the Hubbard model 
has been considered by several groups 
\cite{Mukherjee:2014hsa,Tanizaki:2015rda} 
(see also \cite{Ulybyshev:2019a,Ulybyshev:2019b}
for recent study), 
and the relevance of the contributions from multiple thimbles 
has been reported. 
} 

\section{Sign problem and the (generalized) Lefschetz thimble method
\label{sec:GLTM}}

We start with reviewing Lefschetz thimble methods  
\cite{Cristoforetti:2012su,Cristoforetti:2013wha,
Fujii:2013sra,
Alexandru:2015xva,
Alexandru:2015sua,
Fukuma:2017fjq,
Alexandru:2017oyw,
Fukuma:2019wbv}.

Let $\bbR^N=\{x\}$ be a configuration space 
of $N$-dimensional real variable $x=(x^i)$ $(i=1,\ldots,N)$, 
and $S(x)$ the action. 
Our main concern is to estimate the expectation value 
of an observable $\mathcal{O}(x)$, 
\begin{align}
 \langle \mathcal{O}(x) \rangle_S
 \equiv \frac{
 \int dx\,e^{-S(x)}\,\mathcal{O}(x)
 }{
 \int dx\,e^{-S(x)}
 }. 
\label{vev1}
\end{align}
When $S(x)$ takes complex values ($S(x)=S_R(x) + i\, S_I(x)$), 
one can no longer regard the Boltzmann weight 
$e^{-S(x)}/\int dx\,e^{-S(x)}$ 
as a probability distribution function, 
and thus a direct use of MCMC method 
to estimate \eqref{vev1} is invalidated. 
An obvious workaround is the reweighting with $S_R(x)$: 
\begin{align}
 \langle \mathcal{O}(x) \rangle_{S}
 = \frac{\langle e^{-i\,S_I(x)}\,\mathcal{O}(x) \rangle_{S_R}}{
 \langle e^{-i\,S_I(x)}\rangle_{S_R}}. 
\label{reweighting}
\end{align}
However, 
in MCMC simulations, 
the numerator and the denominator are estimated separately, 
and for large $N$ (or in general, when the action takes large values) 
the integrals of the numerator and the denominator 
may become highly oscillatory 
and take vanishingly small values of $e^{-O(N)}$. 
Then, the estimate of \eqref{reweighting} 
with a sample of size $N_{\rm conf}$
will take the following form:  
\begin{align}
 \langle \mathcal{O}(x) \rangle_{S} \approx
 \frac{e^{-O(N)} + O(1/\sqrt{N_{\rm conf}})}{e^{-O(N)} 
 + O(1/\sqrt{N_{\rm conf}})}.
\end{align}
This means that we need to set the sample size to be exponentially large, 
$N_{\rm conf} = e^{O(N)}$, 
in order to make an estimation with relatively small statistical errors. 
This is the sign problem.

In this article, 
we always assume that both $e^{-S(z)}$ and $e^{-S(z)}\,\mathcal{O}(z)$ 
are entire functions over $\bbC^N$. 
Then, due to Cauchy's theorem in higher dimensions, 
the integrals in \eqref{vev1} do not change under continuous deformations 
of the integration region $\Sigma_t$ ($t\geq 0$) 
with $\Sigma_0=\bbR^N$. 
The sign problem will then get much reduced 
if $\textrm{Im}\, S(z)$ is almost constant 
on some $\Sigma\in \{\Sigma_t\}$. 
In Lefschetz thimble methods, 
such deformations are made according to 
the following antiholomorphic flow equation: 
\begin{align}
 \dot{z}_t^i &= [\partial_i S(z_t)]^\ast,
 \quad
 z^i_{t=0} = x^i. 
\label{flow1}
\end{align}
Equation \eqref{vev1} can then be rewritten as
\begin{align}
 \langle \mathcal{O}(x) \rangle_S 
 = \frac{\int_{\Sigma_t} d  z\, e^{-S(z)}\,\mathcal{O}(z)}
 {\int_{\Sigma_t} d z\, e^{-S(z)}}
 \quad (\Sigma_t \equiv z_t(\bbR^N)), 
\label{LT1}
\end{align}
which can be further rewritten 
as a ratio of reweighted integrals over the parametrization space 
(the same as the original configuration space $\bbR^N$ in this article) 
by using the Jacobian matrix 
$J_t(x)\equiv \bigl(\partial z_t^i(x)/\partial x^j\bigr)$ 
\cite{Alexandru:2015xva}: 
\begin{align}
 \langle \mathcal{O}(x) \rangle_S 
 = \frac{\int_{\bbR^N} d x\, 
 \det\!J_t(x)\,e^{-S(z_t(x))}\,\mathcal{O} (z_t(x))}
 {\int_{\bbR^N} d x\, \det\!J_t(x)\,e^{-S(z_t(x))}} 
 =\frac{
 \bigl\langle e^{i \theta_t(x)} 
 \mathcal{O}(z_t(x))
  \bigr\rangle_{S^{\rm eff}_t}}
 {\bigl\langle e^{i \theta_t(x)} 
 \bigr\rangle_{S^{\rm eff}_t}}. 
\label{LT2}
\end{align}
Here, $S^{\rm eff}_t(x)$ and $\theta_t(x)$ are defined by 
\begin{align}
 e^{-S^{\rm eff}_t(x)} \equiv e^{-{\rm Re}\,S(z_t(x))}\, |\det J_t(x)|,
 \quad
 e^{i \theta_t(x)} \equiv e^{-i\,{\rm Im}\,S(z_t(x))}\, 
 e^{i\, {\rm arg}\, \det J_t(x)},
\end{align}
and $J_t(x)$ obeys the following differential equation 
\cite{Alexandru:2015xva} 
(see also footnote 2 of \cite{Fukuma:2017fjq}): 
\begin{align}
 \dot{J}_t &= [ H(z_t(x))\cdot J_t]^\ast,
 \quad
 J_{t=0} = \pmb{1} 
\label{flow2}
\end{align}
with $H(z)\equiv (\partial_i \partial_j S(z))$. 
In the limit $t\to \infty$, 
$\Sigma_t$ will approach a union of Lefschetz thimbles, 
on each of which ${\rm Im}\,S(z)$ is constant, 
and thus the sign problem is expected to disappear there 
(except for a possible residual and/or global sign problem). 
However, in MCMC calculations 
one cannot take the $t\to\infty$ limit na\"ively, 
because the potential barriers between different thimbles 
become infinitely high. 
This ergodicity problem becomes serious 
when contributions from more than one thimble 
are relevant to estimation.

\section{Tempered Lefschetz thimble method
\label{sec:TLTM}}
In the tempered Lefschetz thimble method (TLTM) \cite{Fukuma:2017fjq}, 
we resolve the dilemma between the sign and ergodicity problems 
by tempering the system with the flow time. 
As a tempering algorithm, 
we adopt parallel tempering \cite{Swendsen1986,Geyer1991} 
because then we need not specify the probability weight factors 
at various flow times 
and because most of relevant steps can be done in parallel processes. 

The algorithm consists of three steps. 
(1) First, we introduce replicas of configuration space, 
each having its own flow time $t_a$ $(a=0,1,\ldots,A)$
with $t_0=0 < t_1 < \cdots < t_A=T$.
Here, the maximal flow time $T$ is chosen 
such that the sign average $|\langle e^{i \theta_T}\rangle|$
is $O(1)$ without tempering. 
(2) We then construct a Markov chain 
that drives the enlarged system 
$(\bbR^N)^{A+1}=\{(x_0,x_1,\ldots,x_A)\}$ 
to global equilibrium 
with the distribution proportional to 
$ \prod_a \exp[-S^{\rm eff}_{t_a}(x_a)].
$
This can be realized 
by combining (a) the Metropolis algorithm 
(or the Hybrid Monte Carlo algorithm \cite{FMU6}) 
in the $x$ direction at each replica 
and (b) the swap of configurations at two adjacent replicas. 
Each of the steps (a) and (b) can be done 
in parallel processes. 
(3) After the system is well relaxed to global equilibrium, 
we estimate the expectation value at flow time $t_a$ 
[see \eqref{LT2}]  
by using the subsample at replica $a$, 
$\{x_a^{(k)}\}_{k=1,2,\ldots,N_{\rm conf}}$, 
that is retrieved from the total sample 
$\{(x_0^{(k)},x_1^{(k)},\ldots,x_A^{(k)})\}_{k=1,2,\ldots,N_{\rm conf}}$: 
\begin{align}
 \frac{
 \bigl\langle e^{i \theta_{t_a}(x)} 
 \mathcal{O}(z_{t_a}(x))
  \bigr\rangle_{S^{\rm eff}_{t_a}}}
 {\bigl\langle e^{i \theta_{t_a}(x)} 
 \bigr\rangle_{S^{\rm eff}_{t_a}}}
 \thickapprox
 \frac{\sum_{k=1}^{N_{\rm conf}}
 \exp[i \theta_{t_a}(x_a^{(k)})]\,
 \mathcal{O} (z_{t_a}(x_a^{(k)}))
  }
 {\sum_{k=1}^{N_{\rm conf}}
 \exp[i \theta_{t_a}(x_a^{(k)})] 
 }
 \equiv \bar{\mathcal{O}}_a.
\label{estimate}
\end{align}
The original proposal in \cite{Fukuma:2017fjq} 
is to use \eqref{estimate} at the maximum flow time, 
$\bar{\mathcal{O}}_{a=A}$, 
as an estimate of $\langle \mathcal{O} \rangle_S$. 
However, the left-hand side of \eqref{estimate} 
is independent of $a$ due to Cauchy's theorem, 
and thus the ratio $\bar{\mathcal{O}}_a$ at large $a$'s 
(where the sign problem is relaxed)
should yield the same value 
within the statistical error margin 
if the system is well in global equilibrium. 
This observation lead us to the following algorithm 
that ensures global equilibrium 
and the sufficiency of the sample size \cite{Fukuma:2019wbv}: 
First, we continue the sampling 
until we find some range of $a$, 
in which 
$\bigl|\overline{e^{i\theta_{t_a}}}\bigr|$ 
are well above $1/\sqrt{2N_{\rm conf}}$ 
(the values for the uniform distribution of phases)
and $\bar{\mathcal{O}}_a$ take the same value 
within the statistical error margin.  
Then, we estimate $\langle \mathcal{O} \rangle_S$ 
by using the $\chi^2$ fit of
$\{\bar{\mathcal{O}}_a\}$ in the region
with a constant function of $a$. 
Global equilibrium and the sufficiency of the sample size 
are checked by looking at the optimized value of 
$\chi^2/{\rm DOF}$.

We here give three comments \cite{Fukuma:2019wbv}. 
(1) In the TLTM, 
one can expect a sufficient overlap between adjacent replicas 
even for large flow times,  
because the distributions at large $a$'s  
have peaks at the same points in $\bbR^N$ 
that flow to critical points in $\bbC^N$. 
(2) The optimal form of $t_a$ is linear in $a$ 
when flowed configurations are close to a critical point, 
because the optimal choice for the overall coefficients 
in tempering algorithms is exponential 
(see, e.g., \cite{Fukuma:2017wzs,Fukuma:2018qgv}) 
and because the real part of the action grows exponentially in flow time 
near critical points. 
See \cite{Fukuma:2019wbv,FMU_Lattice2019_distance} for more details. 
(3) The computational cost in the TLTM 
is expected to be $O(N^{3-4})$. 
Note that the additional increase caused by the tempering algorithm 
(which will be $O(N^{0-1})$) 
can be compensated by the increase of parallel processes.

\section{Application to the Hubbard model away from half filling
\label{sec:Hubbard}}
Let us consider the Hubbard model 
on a $d$-dimensional bipartite lattice with $N_s$ lattice points. 
By using the Trotter decomposition with equal spacing $\epsilon$ 
($\beta = N_\tau \epsilon$), 
and by introducing a Gaussian Hubbard-Stratonovich variable 
$\phi=(\phi_{\ell,\bx})$ $(\ell=0,1,\ldots,N_\tau-1)$, 
the expectation value of the number density operator 
can be expressed in a path-integral form
(see \cite{Fukuma:2019wbv} for the derivation):  
\begin{align}
 \langle n \rangle_S 
 &\equiv \frac{\int [d\phi]\,e^{-S[\phi]}\,n[\phi]}
 {\int [d\phi]\,e^{-S[\phi]}} 
 \quad
 \Bigl( [d\phi] \equiv \prod_{\ell,\bx} d\phi_{\ell,\bx} \Bigr).
\label{n_PI}
\end{align}
Here, 
$e^{-S[\phi]} \equiv e^{-(1/2)\,\sum_{\ell,\bx} \phi_{\ell,\bx}^2}\,
 \det M^a[\phi]\,\det M^b[\phi]
$,  
$M^{a/b}[\phi] 
 \equiv \pmb{1} + e^{\pm \beta \mu}\,\prod_\ell e^{\epsilon \kappa K}\, 
 e^{\pm \,i\sqrt{\epsilon U} \phi_\ell}
$, 
$n[\phi] \equiv $
$(i\sqrt{\epsilon U} N_s)^{-1}\,\sum_\bx \phi_{\ell=0,\bx}
$, 
$\phi_\ell\equiv(\phi_{\ell,\bx}\,\delta_{\bx\by})$, 
and $\prod_\ell$ is a product in descending order. 
Note that the physical quantities depend only on the dimensionless parameters 
$\beta\mu$, $\beta\kappa$, $\beta U$ for fixed $N_\tau$.

\begin{figure}[ht]
\centering
\includegraphics[width=60mm]{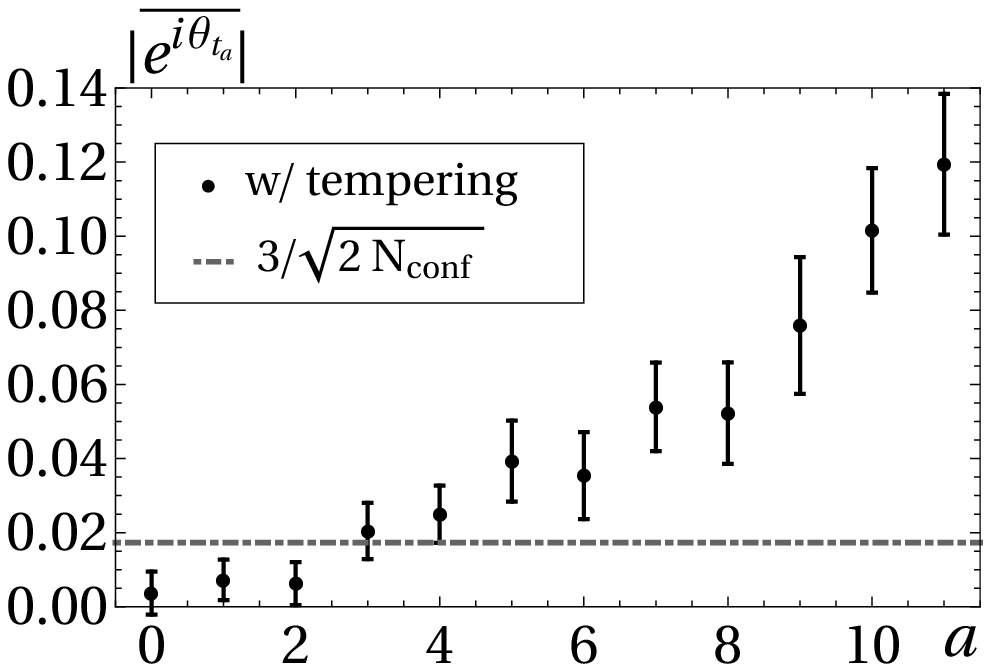} \hspace{10mm}
\includegraphics[width=60mm]{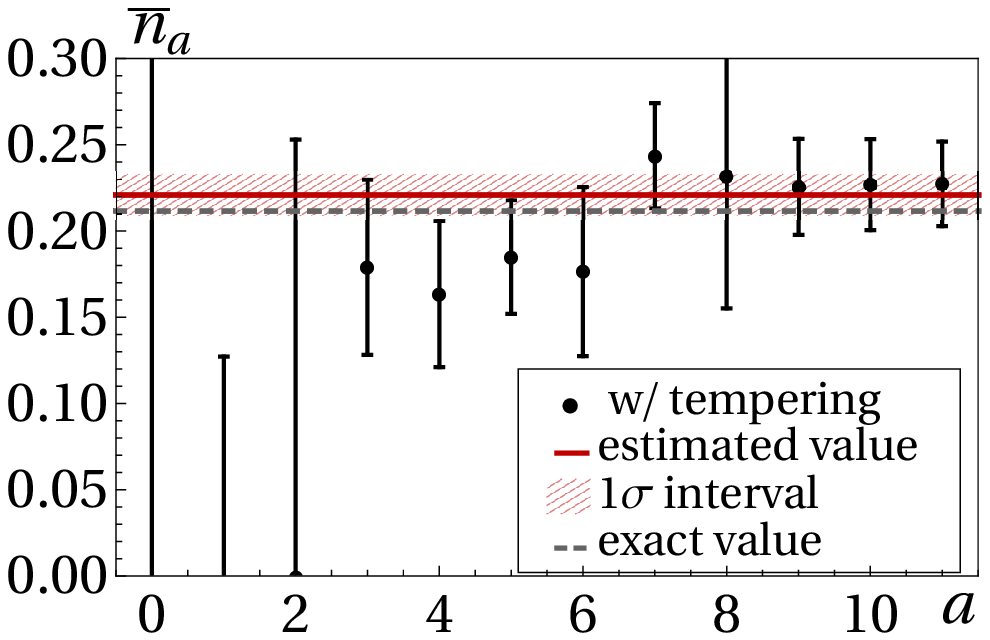}%
\caption{
\label{fig:mu5_tltm}
 With tempering ($\beta\mu=5$) \cite{Fukuma:2019wbv}. 
 (Left) the sign averages 
 at various replicas. 
 The horizontal dashed line represents $3/\sqrt{2N_{\rm conf}}= 0.017$. 
 (Right) the data $\bar{n}_a$. 
 The solid red line with a shaded band represents 
 the estimate of $\langle n \rangle_S$ with $1 \sigma$ interval. 
 The gray dashed line represents the exact value.
}
\end{figure}

\begin{figure}[ht]
\centering
\includegraphics[width=60mm]{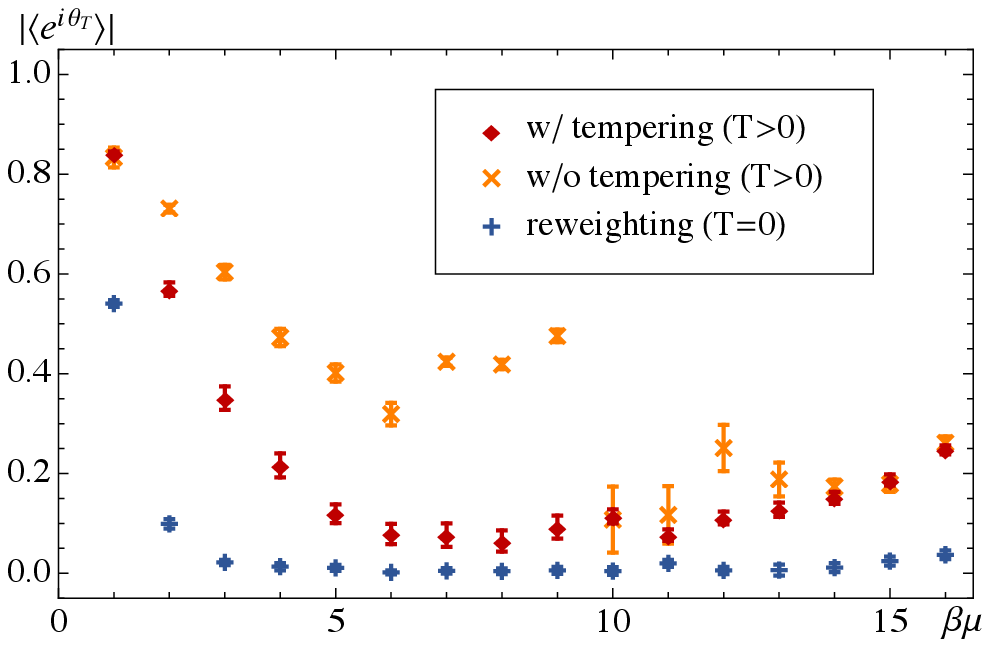} \hspace{10mm}
\includegraphics[width=60mm]{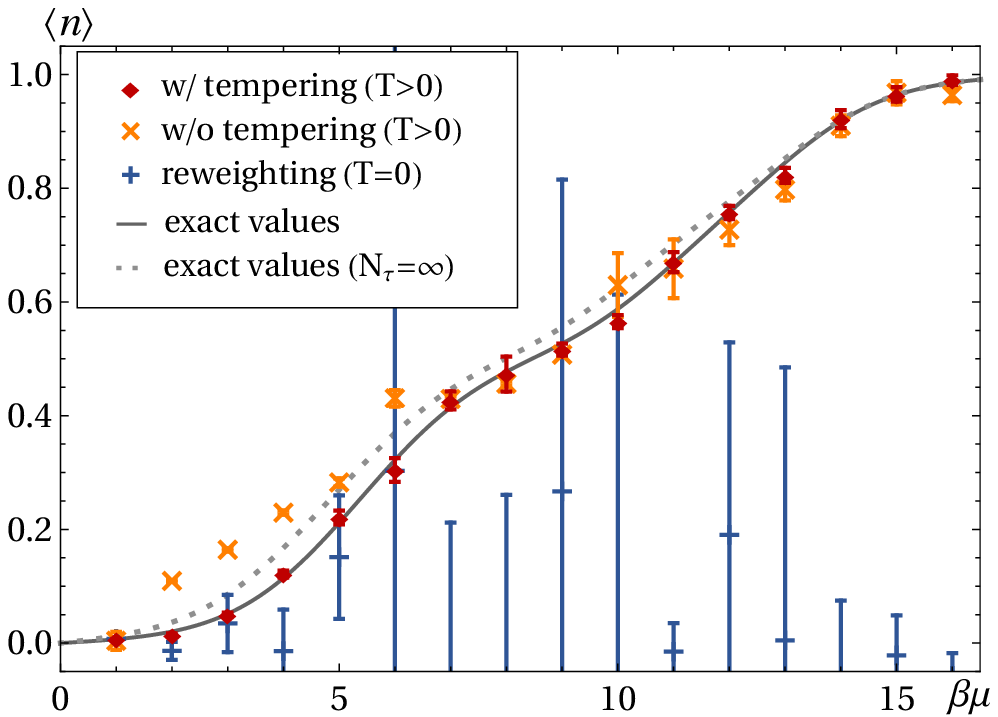}%
\caption{
\label{fig:denom-mu_n-mu}
 (Left) the sign averages at $T$, 
 $|\langle e^{i\theta_T(x)}\rangle_{S^{\rm eff}_T}|$ \cite{Fukuma:2019wbv}.
 (Right) the expectation values of the number density operator, 
 $\langle n \rangle_S$ $(N_\tau=5)$ \cite{Fukuma:2019wbv}. 
 The results obtained with tempering 
 correctly reproduce the exact values. 
 The exact values for $N_\tau=\infty$ are also displayed for comparison. 
}
\end{figure}

\begin{figure}[ht]
\centering
\includegraphics[width=60mm]{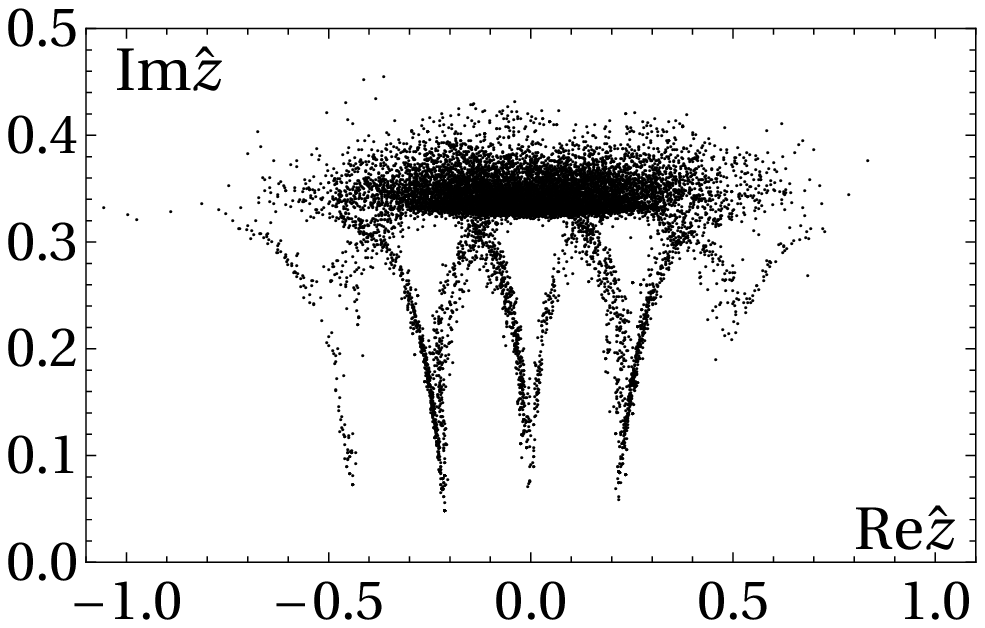} \hspace{10mm}
\includegraphics[width=60mm]{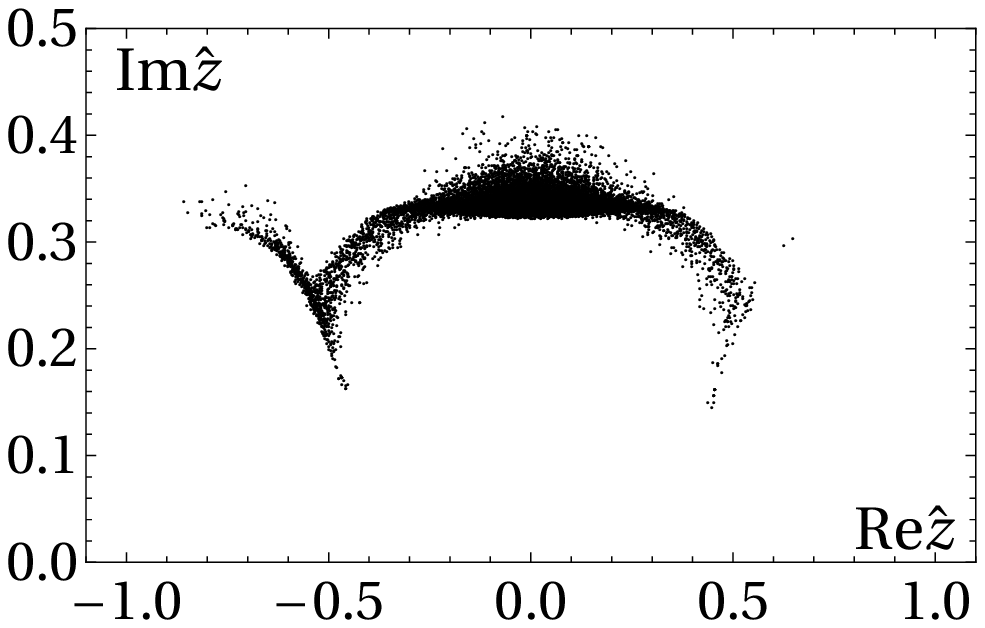}%
\caption{
\label{fig:mu5_a11}
 The distribution of $\hat{z}\equiv (1/N)\sum_i z_T^i$ 
 at $T=0.5$ for $\beta\mu=5$ 
 \cite{Fukuma:2019wbv}. 
 (Left) with tempering. (Right) without tempering.  
}
\end{figure}

We now apply the TLTM to the Hubbard model 
on a two-dimensional periodic square lattice of size $2\times 2$ 
(thus $N_s=4$) 
with $N_\tau=5$. 
We estimate $\langle n \rangle_S$ numerically 
by using the expressions \eqref{n_PI} 
for various values of $\beta\mu$ 
with other parameters fixed to be $\beta\kappa=3$, $\beta U=13$.%
\footnote{ 
  The integral \eqref{n_PI} 
  has a severe sign problem for these parameters 
  as can be seen on the left panel in Fig.~\ref{fig:denom-mu_n-mu}. 
  Note that the extent of the seriousness of the sign problem 
  heavily depends on the choice of the Hubbard-Stratonovich variables, 
  and the present sign problem can actually be avoided 
  within the BSS-QMC method \cite{Blankenbecler:1981jt}. 
  See \cite{Fukuma:2019wbv} for further discussions. 
} 
As an example, 
the sign averages and the data $\{\bar{n}_a\}$ 
for $\beta\mu=5$ with $T=0.5$ 
are shown in Fig.~\ref{fig:mu5_tltm}.  
We see that $\bar{n}_a$ take almost the same values for large $a$'s 
where the sign averages are well above 
the value $1/\sqrt{2 N_{\rm conf}}$. 
The $\chi^2$ fit for $\{a\}=\{5, \ldots, 11\}$ 
gives the estimate $\langle n\rangle_S \approx 0.221 \pm 0.012$ 
(exact value: 0.212)
with $\chi^2/{\rm DOF}=0.45$.
Repeating this analysis for various $\beta\mu$, 
we obtain Fig.~\ref{fig:denom-mu_n-mu} 
\cite{Fukuma:2019wbv}. 
We see that 
the exact values are correctly reproduced 
when the tempering is implemented, 
while there are significant deviations when not implemented. 
As in the $(0+1)$-dimensional massive Thirring model \cite{Fukuma:2017fjq}, 
the deviation reflects the fact that 
the relevant thimbles are not sampled sufficiently. 
This can be explicitly observed 
by looking at the distribution of flowed configurations 
(see Fig.~\ref{fig:mu5_a11}). 

Two comments are in order \cite{Fukuma:2019wbv}. 
(1) A larger value of the sign average does not necessarily mean 
a better resolution of the sign problem 
(see the left panel of Fig.~\ref{fig:denom-mu_n-mu}). 
\begin{figure}[ht]
\centering
\includegraphics[width=60mm]{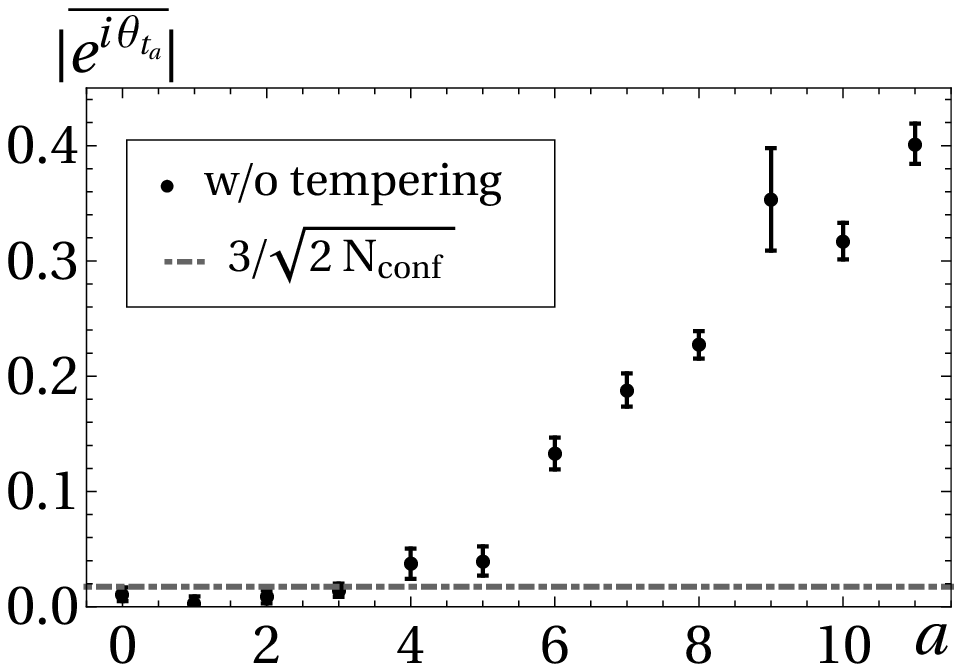} \hspace{10mm}
\includegraphics[width=60mm]{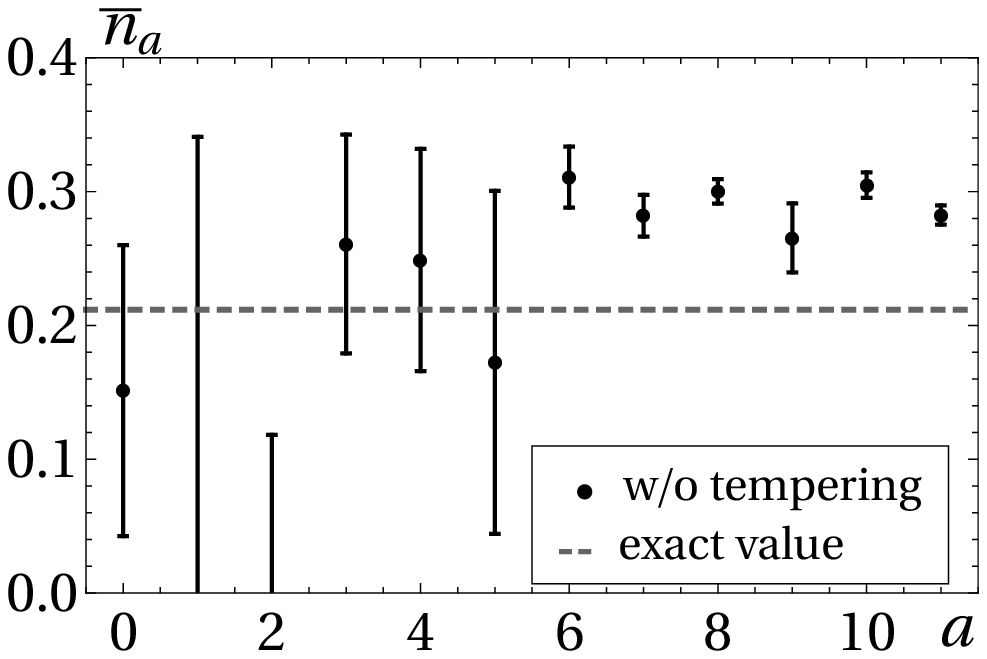}%
\caption{
\label{fig:mu5_gltm}
 Without tempering ($\beta\mu=5$) \cite{Fukuma:2019wbv}. 
 (Left) the sign averages. 
 (Right) the estimates $\bar{n}_a$. 
}
\end{figure}
In fact, when only a very few thimbles are sampled, 
the sign averages can become larger than the correctly sampled values 
due to the absence of phase mixtures among different thimbles. 
(2) From the right panel of Fig.~\ref{fig:mu5_gltm}, we see that
it should be a difficult task 
to find such an intermediate flow time (without tempering) 
that avoids both the sign problem (severe at smaller flow times) 
and the ergodicity  problem (severe at larger flow times. 
See \cite{Fukuma:2019wbv} for more detailed discussions.

\section{Conclusion and outlook
\label{sec:conclusion}}
In this article, 
we reviewed the basics of the TLTM \cite{Fukuma:2017fjq}
and explained a new algorithm \cite{Fukuma:2019wbv}   
which allows a precise estimation within the TLTM. 
We demonstrated its effectiveness 
by applying the TLTM to the two-dimensional Hubbard model 
away from half filling. 
Since the lattice we considered here is still small, 
it must be enlarged much more 
both in the spatial and imaginary time directions 
to claim the validity of our method 
for the sign problem in the Hubbard model, 
revealing its phase structure. 
In doing this, 
it should be important to check 
whether the computational scaling is actually $O(N^{3-4})$ 
as expected. 
More generally, we should keep developing the algorithm further 
so that the TLTM can be more easily applied 
to the three major problems listed in Introduction. 

\acknowledgments
The authors deeply thank the organizers of Lattice 2019. 
They also thank Andrei Alexandru, Yoshimasa Hidaka, Issaku Kanamori, 
Yoshio Kikukawa, Yuto Mori, Jun Nishimura, Akira Ohnishi, Asato Tsuchiya, 
Maksim Ulybyshev, Urs Wenger and Savvas Zafeiropoulos 
for useful discussions and comments. 
This work was partially supported by JSPS KAKENHI 
(Grant Numbers 16K05321, 18J22698 and 17J08709) 
and by SPIRITS 2019 of Kyoto University (PI: M.F.).

\end{document}